\documentclass[12pt]{article} 
\usepackage[table,dvipsnames]{xcolor}
\usepackage[colorlinks, citecolor={blue}]{hyperref}
\usepackage{url}
\usepackage{amsfonts,amscd,amssymb}
\usepackage{amsthm,amsmath,natbib}
\usepackage{algorithm,algorithmicx,algpseudocode}
\usepackage{bm}
\usepackage{bbm} 
\let\oldtabular\tabular
\let\endoldtabular\endtabular
\renewenvironment{tabular}{\rowcolors{4}{white}{trevorblue!15}\oldtabular}{\endoldtabular}
\usepackage{verbatim}
\usepackage{graphicx}
\usepackage{setspace}
\usepackage{natbib}
\usepackage[margin=1in]{geometry}
\usepackage{enumitem}
\usepackage{listings}
\usepackage[textsize=tiny]{todonotes}
\usepackage{tikz}
\usepackage{etoolbox}
\usepackage{appendix}
\usepackage[format=plain,
font=it]{caption}
\usepackage{subcaption}
\usepackage{wrapfig}
\usepackage{xr}
\usepackage{booktabs}
\usepackage{multirow}
\usepackage{authblk}
\usepackage{adjustbox}
\usepackage{floatpag}

\allowdisplaybreaks

\newtoggle{quickdraw}
\toggletrue{quickdraw} 

\definecolor{lightgrey}{rgb}{0.9,0.9,0.9}
\definecolor{darkgreen}{rgb}{0,0.3,0}

\definecolorset{rgb}{}{}{darkred,0.8,0,0;darkgreen,0,0.5,0;darkblue,0,0,0.5}

\definecolor{trevorblue}{rgb}{0.330, 0.484, 0.828}
\definecolor{trevoryellow}{rgb}{0.829, 0.680, 0.306}

\usepackage{color}
\usepackage{fancyvrb}

\DefineVerbatimEnvironment{Highlighting}{Verbatim}{commandchars=\\\{\}}
\usepackage{framed}
\definecolor{shadecolor}{RGB}{241,243,245}
\newenvironment{Shaded}{\begin{snugshade}}{\end{snugshade}}

\newcommand{\AttributeTok}[1]{\textcolor[rgb]{0.40,0.45,0.13}{#1}}

\newcommand{\BuiltInTok}[1]{\textcolor[rgb]{0.00,0.23,0.31}{#1}}

\newcommand{\CommentTok}[1]{\textcolor[rgb]{0.37,0.37,0.37}{#1}}

\newcommand{\ControlFlowTok}[1]{\textcolor[rgb]{0.00,0.23,0.31}{#1}}
\newcommand{\DataTypeTok}[1]{\textcolor[rgb]{0.68,0.00,0.00}{#1}}
\newcommand{\DecValTok}[1]{\textcolor[rgb]{0.68,0.00,0.00}{#1}}

\newcommand{\ExtensionTok}[1]{\textcolor[rgb]{0.00,0.23,0.31}{#1}}
\newcommand{\FloatTok}[1]{\textcolor[rgb]{0.68,0.00,0.00}{#1}}
\newcommand{\FunctionTok}[1]{\textcolor[rgb]{0.28,0.35,0.67}{#1}}
\newcommand{\ImportTok}[1]{\textcolor[rgb]{0.00,0.46,0.62}{#1}}

\newcommand{\KeywordTok}[1]{\textcolor[rgb]{0.00,0.23,0.31}{#1}}
\newcommand{\NormalTok}[1]{\textcolor[rgb]{0.00,0.23,0.31}{#1}}
\newcommand{\OperatorTok}[1]{\textcolor[rgb]{0.37,0.37,0.37}{#1}}

\newcommand{\PreprocessorTok}[1]{\textcolor[rgb]{0.68,0.00,0.00}{#1}}

\newcommand{\StringTok}[1]{\textcolor[rgb]{0.13,0.47,0.30}{#1}}

\title{Scaling Hawkes processes to one million COVID-19 cases}
\date{}

\author[1,2]{Seyoon Ko}
\author[2,3,4]{Marc A.~Suchard}
\author[2]{Andrew J.~Holbrook}
\affil[1]{Department of Mathematics, University of California, Los Angeles}
\affil[2]{Department of Biostatistics, University of California, Los Angeles}
\affil[3]{Department of Biomathematics, University of California, Los Angeles}
\affil[4]{Department of Human Genetics, University of California, Los Angeles}

\graphicspath{{figures/}}

\begin{document}

\maketitle

\begin{abstract}
Hawkes stochastic point process models have emerged as valuable statistical tools for analyzing viral contagion.  The spatiotemporal Hawkes process characterizes the speeds at which viruses spread within human populations.   Unfortunately, likelihood-based inference using these models requires $O(N^2)$ floating-point operations, for $N$ the number of observed cases.  Recent work responds to the Hawkes likelihood's computational burden by developing efficient graphics processing unit (GPU)-based routines that enable Bayesian analysis of tens-of-thousands of observations.  We build on this work and develop a high-performance computing (HPC) strategy that divides 30 Markov chains between 4 GPU nodes, each of which uses multiple GPUs to accelerate its chain's likelihood computations. We use this framework to apply two spatiotemporal Hawkes models to the analysis of one million COVID-19 cases in the United States between March 2020 and June 2023. In addition to brute-force HPC, we advocate for two simple strategies as scalable alternatives to successful approaches proposed for small data settings.  First, we use known county-specific population densities to build a spatially varying triggering kernel in a manner that avoids computationally costly nearest neighbors search.  Second, we use a cut-posterior inference routine that accounts for infections' spatial location uncertainty by iteratively sampling latent locations uniformly within their respective counties of occurrence, thereby avoiding full-blown latent variable inference for 1,000,000 infection locations.  
\end{abstract}


\section{Introduction}\label{sec:intro}

Spatiotemporal Hawkes processes (StHP) are stochastic point processes that model contagion dynamics in space and time  \citep{reinhart2018review}. Applications for this model class include earthquakes \citep{hawkes1973cluster,ogata1988statistical,zhuang2004analyzing,fox2016spatially}, gun violence \citep{loeffler2018gun,park2019investigating,holbrook2021scalable,holbrook2022bayesian}, wildfires \citep{schoenberg2004testing,holbrook2022bayesian} and viral epidemics \citep{kim2011spatio,meyer2014power,choi2015constructing,rizoiu2018sir,kelly2019real,holbrook2022bayesian,holbrook2022viral}.  
Despite the broad applicability of StHP, three significant challenges hinder their application to big data epidemiological studies such as those related to the COVID-19 pandemic.  First, the StHP inherits the computational difficulties shared by most Hawkes process models: likelihood-related computations typically require $O(N^2)$ floating-point operations, where $N$ is the number of observations.  For this reason, application of high-quality StHP models to COVID-19 have restricted themselves to subsets of spatial regions and relatively small periods of time \citep{chiang2022hawkes}.
 Second, StHP rely on spatial data to characterize the probability of future events, so data with low spatial precision can lead to inaccurate inference \citep{holbrook2022bayesian}.   This is especially problematic for epidemiological studies in which spatial data collection is often biased and reporting may purposefully lack precision in the name of privacy.  Third, viruses spread differently in cities, suburbs and rural areas, and regional cultural differences may modulate contagion dynamics.
 When applying StHP to spatiotemporal data from a large geographic region, one must account for spatial inhomogeneities, and one must do this in a scalable manner when $N$ is large.
 In the following, we address these three challenges, taking the large-scale StHP analysis of spatial SARS-CoV-2 contagion as fundamental motivation.

To accomplish this goal, we build on, and deviate from, computational and modeling strategies developed in  \citet{holbrook2021scalable} and \citet{holbrook2022bayesian}.  The former designs parallel computing techniques for the StHP-based analysis of tens-of-thousands of spatiotemporal observations, with implementation using a single GPU.   The latter uses a data augmentation strategy to overcome the problem of spatial coarsening, simultaneously inferring a latent location variable for each observation in the context of a larger Bayesian inference scheme over StHP model parameters.  Unfortunately, these strategies fall short for analyses involving data on the order of one million spatiotemporal observations.  Our solution is twofold.  First, we develop a brute-force approach to address the quadratic computational complexity of the Hawkes model likelihood function (Section \ref{sec:parallel}): we task multiple GPU nodes with generating a single chain each, thereby accomplishing between-chain parallelization, and multiple GPUs accomplish fine-grained parallelization of likelihood evaluations within each of these nodes.  Second, we leverage our high-performance computing strategy in the context of Bayesian cut-posterior inference \citep{plummer2015cuts,jacob2017better} within which we iteratively sample spatial locations directly from their respective prior distributions (Section \ref{sec:cut}).  This strategy eschews the latent variable framework of \citet{holbrook2022bayesian} in order to avoid the impossible task of generating a million-dimensional state space Markov chain.   We combine these two strategies and apply them to two StHP models with spatially homogeneous and inhomogeneous triggering kernels, respectively.  Instead of following \citet{fox2016spatially} and letting an event's distance to its $K$th nearest neighbor parameterize its triggering kernel, we let the kernel's spatial lengscale be proportional to the event county's population density (Section \ref{sec:mods}), thereby avoiding an additional $O(N^2)$ preprocessing step when $N \approx 1{,}000{,}000$.

\section{Methods}\label{sec:meth}

\newcommand{\x}{\mathbf{x}}
\newcommand{\X}{\mathbf{X}}
\newcommand{\dd}{\mbox{d}}
\newcommand{\Id}{\mathbf{I}}
\newcommand{\one}{\boldsymbol{1}}
\newcommand{\ttimes}{\mathbf{t}}
\newcommand{\bbeta}{\boldsymbol{\beta}}
\newcommand{\aalpha}{\boldsymbol{\alpha}}
\renewcommand{\Theta}{\boldsymbol{\theta}}
\newcommand{\z}{\mathbf{z}}

\subsection{Spatiotemporal Hawkes process models}\label{sec:mods}

Consider spatiotemporal data $\{(\x_n,t_n)\}_{n=1}^N$ consisting of positively-valued time points $t_n \in \mathbb{R}_{>0}$  and spatial locations  $\x_n = (x_{n1}, x_{n 2} )\in \mathcal{X}$, for $\mathcal{X} \subset \mathbb{R}^2$ some subset of two-dimensional Euclidean space.    A spatiotemporal Hawkes process \citep{reinhart2018review} models this data as arising from an inhomogeneous Poisson process with instantaneous rate function
\begin{align*}
	\lambda(\x,t) = \mu(t) + \xi(\x,t) = \mu(t) + \sum_{t_n<t} g(\lVert \x - \x_n \rVert, t - t_n) \, ,
\end{align*}
where $\mu$ is an inhomogeneous background rate, $\xi$ is the self-exciting rate component and $g$ is a non-negative triggering function.  In this way, preceding events contribute to the probability of future events occurring within nearby neighborhoods in space and time.  Using the rate function $\lambda$, \citet{daley2003introduction} provide the likelihood function
\begin{align}\label{eq:likelihood}
	\mathcal{L}\left(\{(\x_n,t_n)\}_{n=1}^N\right)
= \exp \left( - \int_{\mathcal{X}} \int_{\mathcal{T}} \lambda(\x,t) \, \dd t\, \dd\x  \right)  \prod_{n=1}^N \lambda(\x_n,t_n) =: e ^{ - \Lambda(\mathcal{X},\mathcal{T}) } \prod_{n=1}^N \lambda_n  \, .
\end{align}
Similar to \citet{holbrook2022bayesian}, we specify background and self-exciting rate components
\begin{align}\label{eq:fixed}
	\mu(t) = \frac{\mu_0}{A(\mathcal{X})} \sum_{n=1}^N \mathcal{I}_{[t\neq t_n]}  \phi_1 (t|t_n, \tau_t)  \, , \quad	\xi_1(\x,t) = \frac{\xi_0}{\sigma_t} \sum_{t_n<t}  e^{-  (t-t_n)/ \sigma_t }  \phi_2( \x | \x_n,\sigma_x) \, ,
\end{align}
where $\mathcal{I}$ is the indicator function, $A(\mathcal{X})$ is the area of the spatial domain, $\phi_1$ is a $1$-dimensional Gaussian density, $\phi_2$ is a $2$-dimensional spherical Gaussian density, and $\mu_0$, $\tau_t$, $\sigma_x$, $\sigma_t$, and $\xi_0$ are strictly positive parameters.  We call $\tau_t$ the background temporal lengthscale and $\sigma_x$ ($\sigma_t$) the triggering kernel's spatial (temporal) lengthscale.  \citet{rizoiu2018sir} relate $\xi_0$ to the basic reproduction number $R_0$ of SIR-type compartmental models, but the two quantities are only exactly equal for infinite-dimensional populations or at the very beginning of an epidemic.    Since the analysis of Section \ref{sec:data} does not meet these conditions, we do not use the symbol $R_0$.  We do, however, note that a higher $\xi_0$ increases the probability that an event triggers subsequent events.

Following \citet{fox2016spatially}, we also consider a second self-exciting component model $\xi_2$ that allows for a spatially-varying spatial kernel.  \citet{fox2016spatially} build such a kernel using observation-specific spatial lengthscales that involve the spatial radius of an observations circular neighborhood that envelopes its $K$ nearest neighbors, but this approach becomes computationally onerous in big data contexts.  Here, we use local population densities $D_n$ to modulate the spatially-varying kernel instead:
 \begin{align}\label{eq:varying}
 \xi_2(\x,t) = \frac{\xi_0}{\sigma_t} \sum_{t_n<t}  e^{-  (t-t_n)/ \sigma_t }  \phi_2\left( \x | \x_n,\sigma_x/\sqrt{D_n}\right) \, .
 \end{align}
 In the SARS-CoV-2 data analysis of Section \ref{sec:data}, for example, $D_n$ is the population density measured as persons per square-mile for the U.S.~county in which event $n$ occurs.  
 
 Instead of dealing with the exact form of the likelihood \eqref{eq:likelihood} for each of these models, we use the convenient approximation $\mathcal{X} \approx \mathbb{R}^2$ for the triggering kernel.  Separability of spatial and temporal kernels combined with normalization of the latter then lead to the approximate likelihood 
 \begin{align*}
 	\mathcal{L}\left(\{(\x_n,t_n)\}_{n=1}^N\right) \approx e ^{ - \Lambda(\mathcal{T}) } \prod_{n=1}^N \lambda_n  \, ,  
 \end{align*}
where $\Lambda(\mathcal{T})$ is the integral of the weighted temporal kernels over the temporal domain. We let $\mathcal{T}= (0,t_N]$ and follow \citet[Section 3.2]{laub2015hawkes} to obtain
\begin{align*}
	\Lambda(\mathcal{T}) &= \mu_0 \sum_{n=1}^N \left(\Phi\left(\frac{t_N-t_n}{\tau_t} \right) -\Phi\left(\frac{-t_n}{\tau_t} \right) \right) - \xi_0 \sum_{n=1}^N \left( e^{-(t_N-t_n)/\sigma_t} -1 \right)  \\ \nonumber
	&= \sum_{n=1}^N  \left[ \mu_0\left(  \Phi\left(\frac{t_N-t_n}{\tau_t} \right) -\Phi\left(\frac{-t_n}{\tau_t} \right)\right)- \xi_0 \left( e^{-(t_N-t_n)/\sigma_t} -1 \right) \right]  =: \sum_{n=1}^N \Lambda_n \, .
\end{align*}
Here, $\Phi$ is the standard normal cumulative density function. Letting $\Theta$ denote the totality of model parameters, the resulting log-likelihood for the first model is
\begin{gather}\label{eqn:model}
	\ell(\{(\x_n,t_n)\}_{n=1}^N|\Theta) = - \Lambda(\mathcal{T}) + \sum_{n=1}^N\log \lambda_n   \\ \nonumber
	= \sum_{n=1}^N \left[ \log \left(  \sum_{n'=1}^N \frac{\mu_0\mathcal{I}_{[t_{n'}\neq t_n]} }{A(\mathcal{X})} \phi_1 (t_n |t_{n'}, \tau_t)
	+\frac{\xi_0  \mathcal{I}_{[t_{n'}<t_n]}}{ \sigma_t }  e^{-  (t_n-t_{n'})/ \sigma_t }  \phi_2( \x_n | \x_{n'},\sigma_x) \right)   - \Lambda_n \right] \\ \nonumber
	=: \sum_{n=1}^N \left[
	\log \left(  \sum_{n'=1}^N \lambda_{nn'} \right)  - \Lambda_n \right] =: \sum_{n=1}^N \ell_n  \, .
\end{gather}
The self-exciting component with spatially-varying kernel leads to an analogously structured log-likelihood.  Importantly, this structure and its double summation over indices $n$ and $n'$ leads to $O(N^2)$ time complexity for each likelihood evaluation.   We address this inferential bottleneck in Section \ref{sec:parallel} after establishing the larger inference context in Section \ref{sec:cut}.

\subsection{Cut-posterior inference}\label{sec:cut}

\newcommand{\obsData}{\mathfrak{x}}
\renewcommand{\t}{\mathbf{t}}
\newcommand{\ObsData}{\mathfrak{X}}

We would like to apply the Hawkes process models of Section \ref{sec:mods} to spatiotemporal events with spatial coordinates $\x_n$ that are arbitrarily valued within a given domain.   Unfortunately, we only observe coarsened spatial data $\obsData_n$ \citep{heitjan1991ignorability,heitjan1993ignorability}.  Examples of coarsening include rounding, truncation and censoring.  \citet[Section 2.2]{holbrook2022bayesian} discuss spatial coarsening and develop data augmentation schemes that account for spatial coarsening within the analysis of three real-world data examples.  Assuming a stochastic coarsening mechanism is ignorable in the sense of \citet{heitjan1991ignorability}, \citet{holbrook2022bayesian} place prior distributions $p(\x_n)$ over unobserved locations $\x_n$ and infer these latent variables along with model parameters $\Theta$ using MCMC to target their joint posterior distribution
\begin{align}\label{eq:post}
	p(\Theta, \X | \ObsData , \t)  =	\frac{\mathcal{L}\left(\X,\t | \Theta \right) p(\Theta)  p(\ObsData|\X) p(\X)}{\int \int \mathcal{L}\left(\X,\t | \Theta \right) p(\Theta)  p(\ObsData|\X) p(\X) d\X d\Theta}  = 	\frac{\mathcal{L}\left(\X,\t | \Theta \right) p(\Theta)  p(\ObsData|\X) p(\X)}{p(\ObsData,\t)} \, ,
\end{align}
where we use the notation $\X=\{\x_n\}_{n=1}^N$, $\t=\{t_n\}_{n=1}^N$ and $\ObsData=\{\obsData_n\}_{n=1}^N$, letting $p(\Theta)$ be the prior specified over model parameters $\Theta$, $p(\ObsData|\X)$ be the probability mass or density function describing the stochastic spatial coarsening mechanism and $p(\X)=\prod_{n=1}^N p(\x_n)$.   Using MCMC to generate samples from \eqref{eq:post} is not difficult when $N$ is in the low thousands, but the extreme dimensionality $D=2N+|\Theta|$ of the MCMC state space and the resulting autocorrelation betweeen samples make the computational task infeasible when $N$ approaches the scale we consider in Section \ref{sec:data}.   

We avoid the extreme autocorrelation associated with having a million-dimensional state space by instead sampling from the cut-posterior distribution \citep{plummer2015cuts}:
\begin{align}\label{eq:cut1}
p_{\mbox{c}}(\Theta, \X | \ObsData,\t)  &=  p(\Theta | \X, \t) \, p(\X|\ObsData) \\
&= \frac{p(\ObsData,\t)}{p(\ObsData) p(\t|\X)}  p(\Theta, \X | \t,\ObsData) \, . \label{eq:cut2}
\end{align} 
When $\X$ is a nuisance parameter, one calls $p(\t|\X)=\int p(\t|\X,\Theta) p(\X|\Theta) p(\Theta) d\Theta$ the `feedback' term, inclusion of which in the denominator of \eqref{eq:cut2} allows one to `cut' feedback or remove the influence of $\t$ on $\X$ \citep{jacob2017better}.   On the other hand, the form of \eqref{eq:cut1} suggests that one may use MCMC to generate samples from $p_{\mbox{c}}$ by iterating between sampling from $p(\X|\ObsData)$ and any MCMC kernel that leaves $p(\Theta|\X,\t)$ invariant.  In the data analysis of Section \ref{sec:data}, we use the factored form $p(\X|\ObsData) \propto \prod_{n=1}^N p(\obsData_n|\x_n) p(\x_n)$ to update $\X$ in time $O(N)$; conditioned on $\X$, we use the high-performance computing strategy described in Section \ref{sec:parallel} to parallelize the Hawkes likelihood computations involved in a univariate-update Metropolis-Hastings kernel that maintains detailed balance with respect to $p(\Theta|\X,\t)$. 

\subsection{Software implementation}\label{sec:parallel}

\subsubsection{Kernel improvement}
We write our GPU code using the Open Computing Language (OpenCL). Within this framework, one writes functions called kernels, which the library compiles at runtime and assigns to individual work groups of GPU cores separately for parallel execution.  We rewrite the OpenCL kernel of the hpHawkes R package \citep{holbrook2021scalable} to improve numerical stability. With the previous kernel, the log-likelihood computation often faces numerical instability, resulting in not a number (\texttt{NaN}) values and making it impossible to reliably execute MCMC in single precision, which is often much faster than double precision on GPUs. Even with double precision, numerical instability sometimes arises for the hpHawkes kernel when $N$ approaches one million. We address this issue in two ways. First, we apply the log-sum-exp trick (see, for example, Sec. 5 of \citet{murphy2006naive}) for the inner summation within the log function of \eqref{eqn:model}. Second, we replace the values of $\sum_{n'=1}^N \lambda_{nn'}$ smaller than $10^{-40}$ with $10^{-40}$ before taking the logarithm. The latter changes the log-likelihood value, but is only applied for low-likelihood regions.  Appendix \ref{sec:kernel} contains the kernel responsible for computing the log-likelihood \eqref{eqn:model} in single precision as an example.

\subsubsection{Multi-GPU computation}
With the improved kernel, we distribute the computation over multiple GPUs. The heavy $O(N^2)$ computation of \eqref{eqn:model} divides across GPUs, uniformly partitioning $N$ terms across $G$ GPUs allocated for the chain. Each GPU $g$ computes a slice $\sum_{n=b_g}^{e_g} \ell_n$, where $b_g$ and $e_g$ indicating the beginning and ending index of sample points allocated for GPU $g$. We then use Message Passing Interface (MPI, \citet{mpi40}) to compute the total likelihood by gathering the partial sums to one process. Multiple processes launch, each of which uses one CPU core to drive one GPU. We run within-chain parallelization mainly on multiple GPUs on a single machine, but the software design allows the utilization of GPUs on different computing nodes if they are interconnected.
Also, multiple MCMC chains execute as separate jobs on a computing cluster with all $M$ GPU nodes available managed by a job scheduler, thus utilizing most of the available GPUs in the cluster simultaneously.

\subsubsection{Julia}
We use the Julia programming language \citep{bezanson2017julia} for software implementation. Julia is a high-level programming language with fast runtime targeting scientific computing with just-in-time compilation. We use the OpenCL interface in Julia\footnote{\url{https://github.com/JuliaGPU/OpenCL.jl}} due to its simpler configuration and reduced need for wrapper code.
The outline of the implementaion is as follows. First, we define a Julia function returning the OpenCL kernel in the form of a string. Using metaprogramming, we can automatically create kernels for different sets of types, such as single precision and double precision. We have a kernel for computing contributions to the log-likelihood contribution of each observation $n$ and a kernel for computing a sum of elements in an array.  
Second, OpenCL's program build interface compiles and builds each kernel into executable form. 
Next, a driver function calls and implements the kernel functions to return the log-likelihood. Finally, the MCMC function is implemented using the driver functions.
We define a Julia function \texttt{lik\_contribs\_kernel\_model1} that returns the string containing the kernel functions for single precision or double precision, depending on arguments. Then, the kernel programs are built using the following function:
\begin{Shaded}
\begin{Highlighting}[numbers=left,fontsize=\footnotesize,]
\ImportTok{using} \BuiltInTok{OpenCL}\NormalTok{, }\BuiltInTok{Pipe}
\KeywordTok{function} \FunctionTok{get\_kernels}\NormalTok{(ctx}\OperatorTok{::}\DataTypeTok{cl.Context}\NormalTok{, T}\OperatorTok{::}\DataTypeTok{Type\{\textless{}:AbstractFloat\}}\NormalTok{)}

    \CommentTok{\# define OpenCL program and kernel}
\NormalTok{    p\_loglik1\_cb }\OperatorTok{=} \PreprocessorTok{@pipe}\NormalTok{ cl.}\FunctionTok{Program}\NormalTok{(ctx, source}\OperatorTok{=}
        \FunctionTok{lik\_contribs\_kernel\_model1}\NormalTok{(}\FunctionTok{Val}\NormalTok{(T))) }\OperatorTok{|\textgreater{}}
        \NormalTok{cl.}\FunctionTok{build!}\NormalTok{(\_; options}\OperatorTok{=}\StringTok{"{-}cl{-}fast{-}relaxed{-}math"}\NormalTok{)}
\NormalTok{    k\_loglik1\_cb }\OperatorTok{=}\NormalTok{ cl.}\FunctionTok{Kernel}\NormalTok{(p\_loglik1\_cb, }\StringTok{"computeLikContribsModel1"}\NormalTok{)}
    \ControlFlowTok{return} \NormalTok{k\_loglik1\_cb}
\KeywordTok{end}
\end{Highlighting}
\end{Shaded}
The driver function below queues the kernels with appropriate arguments, including Hawkes model parameters \texttt{sigmaXprec}, \texttt{tauTprec}, \texttt{omega},  \texttt{theta} and \texttt{mu0}. The struct \texttt{HawkesStorage\{T\}} contains the kernel program and temporary buffers. The kernel \texttt{k\_sum} computes the sum of likelihood contributions for each observation $n$.
\begin{Shaded}
\begin{Highlighting}[numbers=left,fontsize=\footnotesize,]
\NormalTok{const TPB  = 256}
\KeywordTok{function} \FunctionTok{loglik1\_cb}\NormalTok{(d}\OperatorTok{::}\DataTypeTok{HawkesStorage\{T\}}\NormalTok{,}
\NormalTok{    sigmaXprec}\OperatorTok{::}\DataTypeTok{T}\NormalTok{, tauXprec}\OperatorTok{::}\DataTypeTok{T}\NormalTok{, tauTprec}\OperatorTok{::}\DataTypeTok{T}\NormalTok{, omega}\OperatorTok{::}\DataTypeTok{T}\NormalTok{, theta}\OperatorTok{::}\DataTypeTok{T}\NormalTok{, mu0}\OperatorTok{::}\DataTypeTok{T}\NormalTok{, 2;}
\NormalTok{    first\_idx }\OperatorTok{=} \FloatTok{0}\NormalTok{,}
\NormalTok{    last\_idx }\OperatorTok{=}\NormalTok{ d.m}
\NormalTok{    ) }\KeywordTok{where}\NormalTok{ T}
\NormalTok{    d.}\FunctionTok{queue}\NormalTok{(d.k\_loglik1\_cb, (last\_idx }\OperatorTok{{-}}\NormalTok{ first\_idx) }\OperatorTok{*}\NormalTok{ TPB, TPB,}
    \NormalTok{d.locations\_buff, d.times\_buff, d.likContribs\_buff, }
\NormalTok{        sigmaXprec, tauXprec, tauTprec, omega, theta, mu0, }
        \FunctionTok{Int32}\NormalTok{(dimX), }\FunctionTok{UInt32}\NormalTok{(first\_idx), }\FunctionTok{UInt32}\NormalTok{(last\_idx), }\FunctionTok{UInt32}\NormalTok{(d.m)) }\OperatorTok{|\textgreater{}}\NormalTok{ cl.wait}
\NormalTok{    cl.}\FunctionTok{copy!}\NormalTok{(d.queue, d.likContribs, d.likContribs\_buff) }\OperatorTok{|\textgreater{}}\NormalTok{ cl.wait}
\NormalTok{    d.}\FunctionTok{queue}\NormalTok{(d.k\_sum, TPB, TPB, d.likContribs\_buff, d.likContribs\_reduced\_buff, }
        \FunctionTok{UInt32}\NormalTok{((last\_idx }\OperatorTok{{-}}\NormalTok{ first\_idx) }\OperatorTok{÷} \FloatTok{8}\NormalTok{)) }\OperatorTok{|\textgreater{}}\NormalTok{ cl.wait}
\NormalTok{    cl.}\FunctionTok{copy!}\NormalTok{(d.queue, d.likContribs\_reduced, d.likContribs\_reduced\_buff) }\OperatorTok{|\textgreater{}}\NormalTok{ cl.wait}
    \FunctionTok{sum}\NormalTok{(d.likContribs\_reduced)}
\KeywordTok{end}
\end{Highlighting}
\end{Shaded}
We use this function inside a full MCMC function that implements Metropolis-Hastings with univariate updates and  cut-posterior inference. The updates of locations $\X$ within this cut-posterior inference during the analysis of Section \ref{sec:data} use Julia packages GeoStats and GeoIO \citep{Hoffimann2023}, which sample data points uniformly from regions with polygonal boundaries.  We collect these functions within the HPHawkes.jl Julia package (\url{https://github.com/kose-y/HPHawkes.jl}). 

\section{Results}

\subsection{Computational performance}\label{sec:perform}

\begin{figure}[!t]
\begin{center}
 \includegraphics[width=0.45\textwidth]{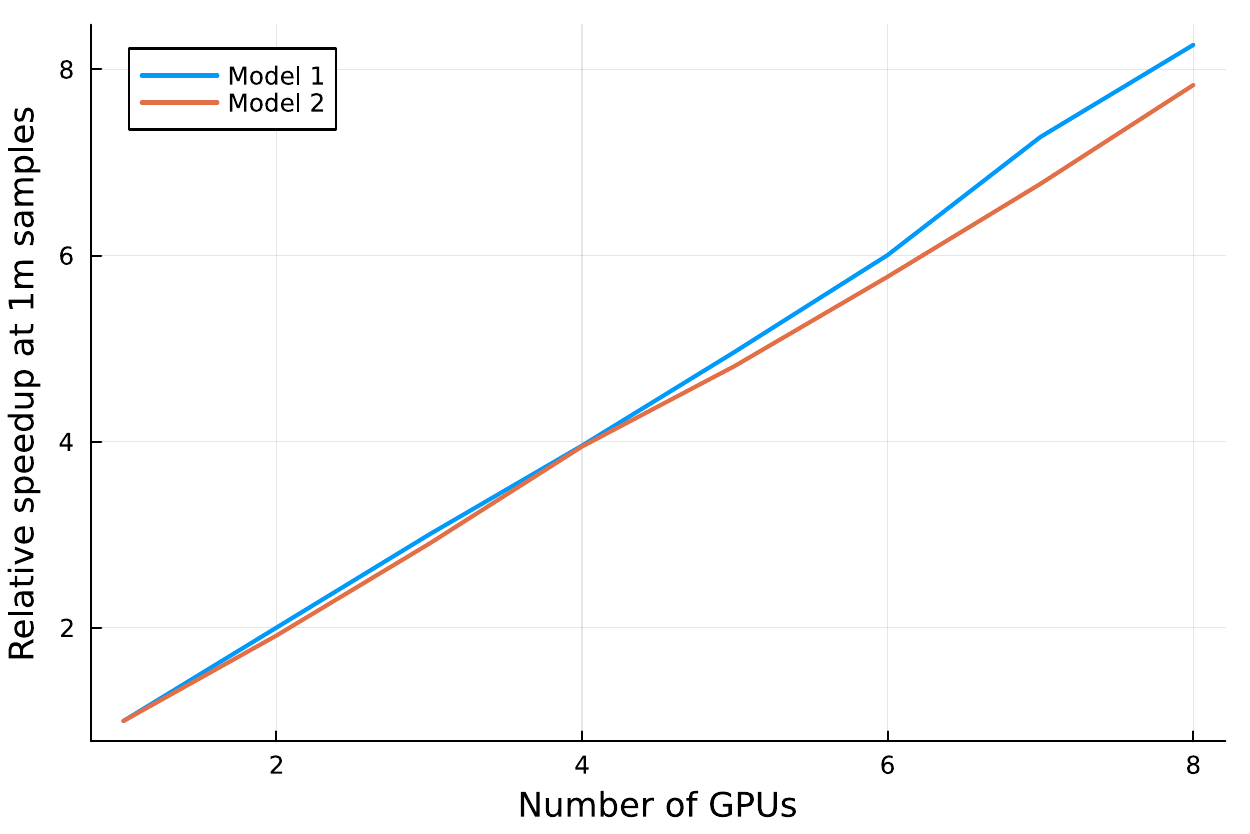} 
 \end{center}
\begin{center}
\includegraphics[width=0.45\textwidth]{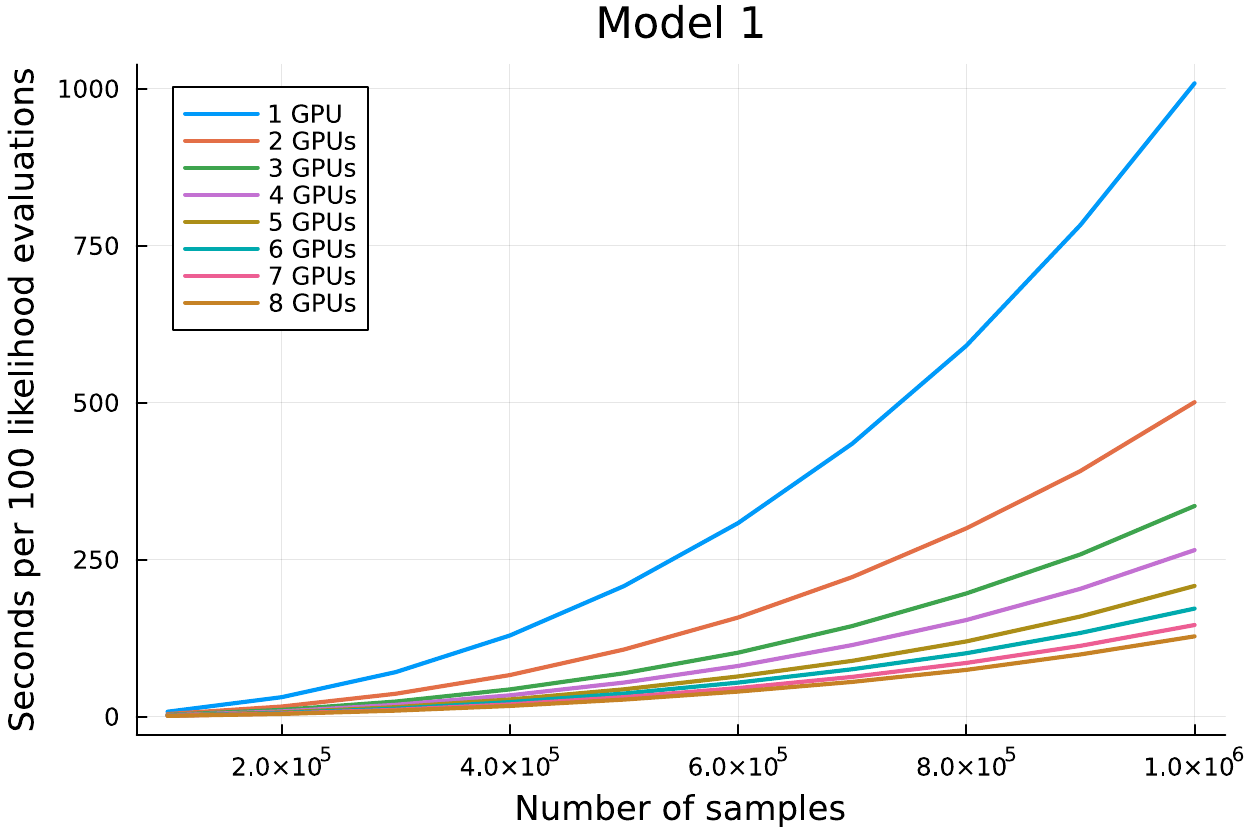}
\includegraphics[width=0.45\textwidth]{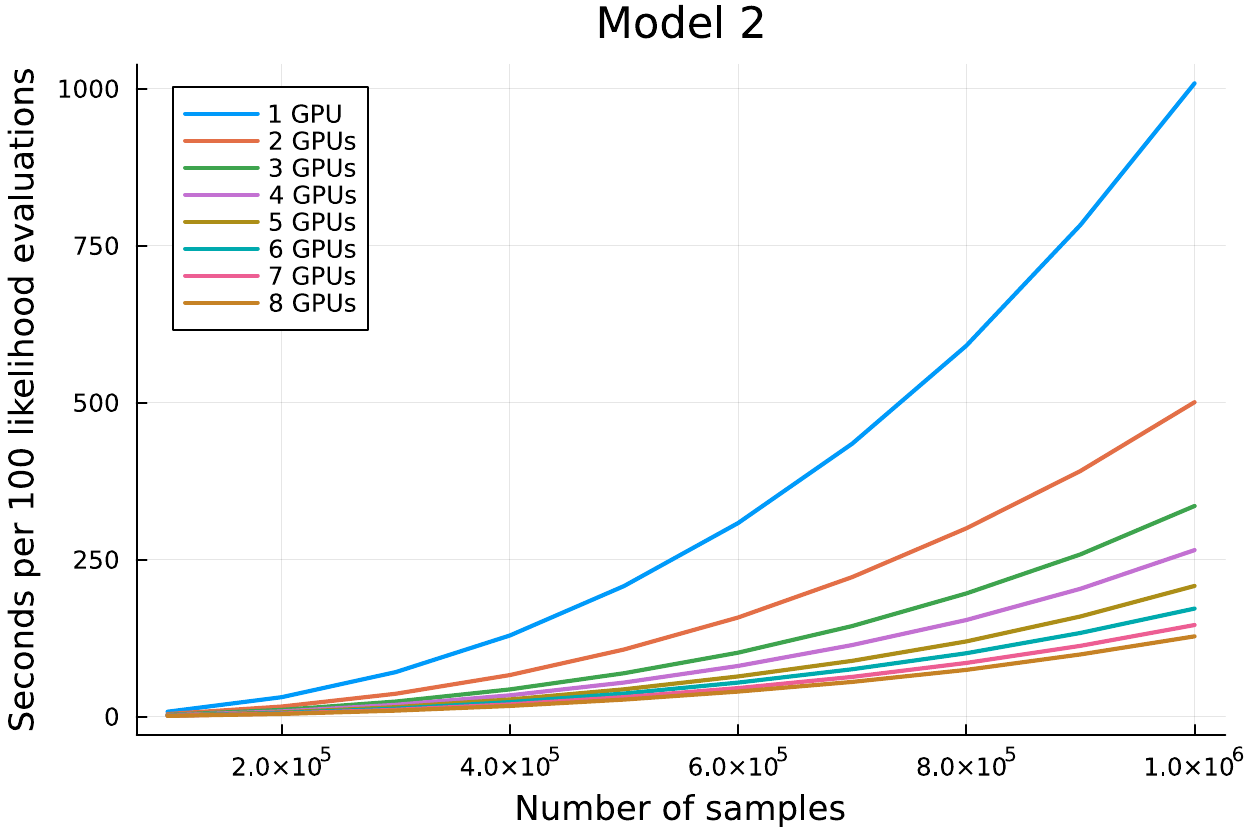}
\end{center}
\caption{Timing for spatiotemporal Hawkes process log-likelihood computations. (Top) Relative speedups over single-GPU run conditioning on one million observations.  The single-GPU implementation is 1.7-times faster than that of \citep{holbrook2021scalable} for double-precision, which itself is over 200-times faster than a single-core C++ implementation. (Bottom) Seconds per 100 log-likelihood evaluations from 100,000 to one million observations for Model 1 (left) and Model 2 (right). Model 2, which has a spatially-varying triggering kernel, incurs little extra computational cost compared to Model 1, which has a spatially-constant kernel.}\label{fig:speedup}
\end{figure}

With optimized GPU kernels and the distribution of computation over multiple GPUs, we achieve numerical stability and a significant overall speedup. 
For the GPU results, we use up to 32 AMD Radeon Instinct MI50 GPUs, each with 3840 cores and 16GB of memory. Eight GPUs are installed on a single node with an AMD EPYC 7642 processor, which has 48 CPU cores. We use up to four nodes of the same configuration.

Using the previous hpHawkes kernel of \citep{holbrook2021scalable}, it is impossible to reliably perform the analysis with over 100,000 observations with single precision and over 1,000,000 observations with double precision due to numerical instability. By clipping the log-likelihood at a small positive number, the kernel now avoids numerical instability with 1,000,000 data points on single precision. Thanks to additional optimization, the double-precision kernel is now 1.7 times faster than before, and we can reliably run the single-precision kernel 7.2 times faster than the previous double-precision kernel with 1,000,000 observations.
The use of multiple GPUs also scales up the MCMC run of each chain. Due to the nature of computation with low memory bandwidth and many GPU cores, the speed increase is nearly proportional to the number of GPUs used, up to 8 GPUs installed within the same node, as shown in the top plot of Figure \ref{fig:speedup}.
The likelihood formula suggests that the computation time increases quadratically with $N$. We run 100 likelihood computations for different $N$ values with varying numbers of GPUs, and we observe the expected quadratic trend in the bottom plots of Figures \ref{fig:speedup}. The parallelization does not change the asymptotic complexity, but it reduces the required computational time.


\subsection{Analyzing SARS-CoV-2 contagion}\label{sec:data}

\begin{figure}[!t]
	\centering
	\includegraphics[width=0.9\textwidth]{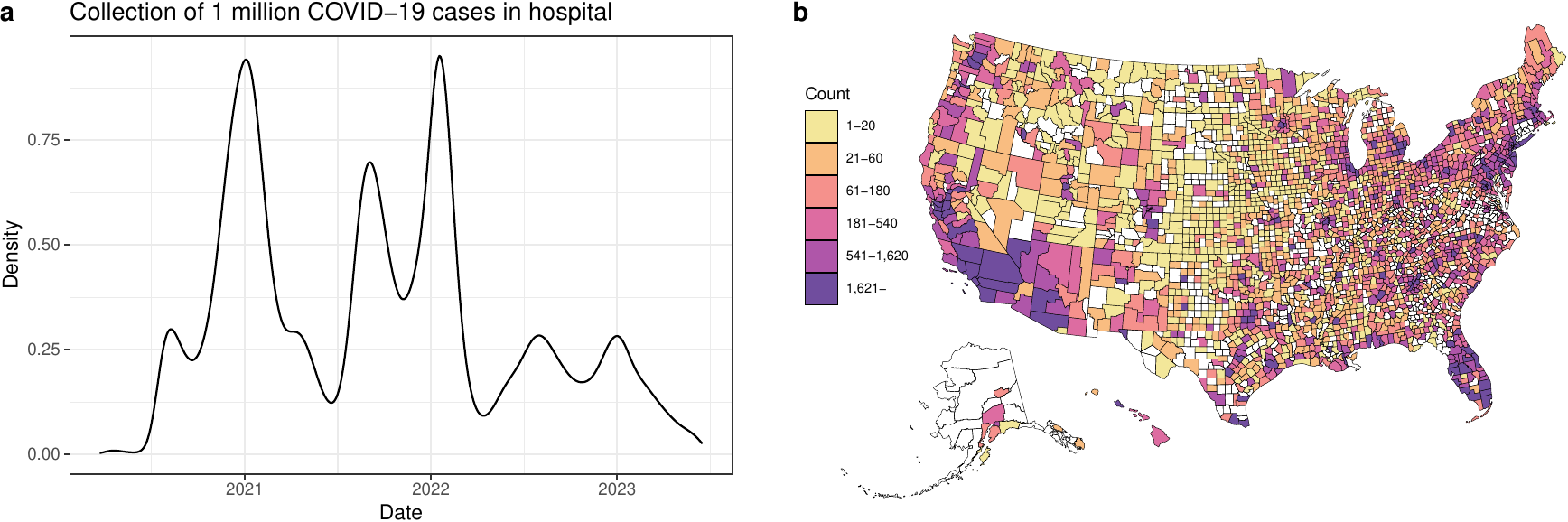}
	\caption{Distribution of one million COVID-19 cases detected in U.S.~hospitals between March 2020 and June 2023.}\label{fig:data}
\end{figure}

We apply our framework to the analysis of 1,000,000 confirmed COVID-19 cases in U.S.~hospitals between March 2020 and June 2023, using the COVID-19 Reported Patient Impact and Hospital Capacity by Facility data obtained in June 2023 from \url{www.healthdata.gov}.  The data aggregates weekly cases and features double counting: a week's case is also ascribed to subsequent weeks if the patient remains in hospital.  We randomly assign cases to days within their ascribed week and down-sample the resulting data (\url{https://doi.org/10.5281/zenodo.12735370}) to get our final dataset consisting of 1 million cases (Figure \ref{fig:data}).   Conditioning on this data, we use MCMC to sample from the cut-posterior distribution \eqref{eq:cut1} for both the StHP model with fixed triggering kernel \eqref{eq:fixed} and the model with spatially-varying kernel \eqref{eq:varying}.  For both models, the MCMC routine iterates between updating the cut distribution's factors  $p(\Theta|\X,\t)$ (the distribution of model parameters given event locations and times) and $p(\X|\ObsData)$ (the distribution of event locations given their hospitals' counties).  For each model, we use the software setup described in Section \ref{sec:parallel} and the 4 GPU nodes described in Section \ref{sec:perform} to generate 30 chains for 20,000 iterations each.  We assign two chains at a time to each GPU node and its 8 GPUs, using 4 GPUs to accelerate each chain's respective likelihood calculations.  In this manner, individual chains for the spatially-constant model \eqref{eq:fixed} require less than 18 hours to complete, and all 30 chains for the same model require roughly 3 days.  The same numbers are 23 hours and 4 days for the spatially-varying model \eqref{eq:varying}.

After removing the first 1,000 MCMC iterations as burn-in, we use bulk and tail effective sample size (ESS) and R-hat to measure sample quality for each model parameter \citep{vehtari2021rank_}.   For the model with spatially-fixed triggering kernel, the spatial lengthscale parameter $\sigma_x$ mixes most poorly, having the largest R-hat (1.006), smallest bulk ESS (5,926.23) and smallest tail ESS (10,426.33) of all model parameters.
These values suggest more than adequate convergence: the R-hat is well below the prescribed upper bound of 1.05, and the two ESS numbers are large.  The spatially-varying model exhibits marginally better mixing.  Again, the spatial lengthscale parameter mixes most poorly, having the largest R-hat (1.005), and smallest bulk ESS (6,300.04) and tail ESS (11,041.02) of all parameters.  We use the RStan package \citep{rstan} written for the R language \citep{rlang} to compute these diagnostics.
We obtain empirical posterior medians and 95\% credible intervals based on these samples and present these posterior summaries in Table \ref{tab:medians}.  Both models' posteriors for the self-excitatory weight $\xi_0$ are similar, and the ratio $\xi_0/(\xi_0+\mu_0)$, which one can interpret as the proportion of events the model believes to arise from local contagion, has a posterior median of 0.998 for the spatially-constant model and a marginally increased 0.999 for the spatially-varying model.  The two models' temporal lengthscales $\sigma_t$, which indicate the expected delay between a case and its preceding case, have statistically (but perhaps not practically) significantly different posteriors: the posterior median for the spatially-constant model is 2.81 weeks (95\% CrI: 2.79, 2.83) compared to 2.50 weeks (2.49, 2.52) for the spatially-varying model. 

Table \ref{tab:lengthscales} shows posterior medians and 95\% credible intervals of effective spatial lengthscales in miles for a number of U.S.~counties.  For the spatially-constant triggering function model \eqref{eq:fixed}, the effective spatial lengthscale is exactly $\sigma_x$; for the spatially-varying triggering function model \eqref{eq:varying}, the effective spatial lengthscales are $\sigma_x / \sqrt{D_n}$, where $D_n$ is the population density of the county in which event $n$ occurs, measured in persons per square-mile.  For both models, we interpret the effective spatial lengthscale as the expected distance between a case and its preceding case after time $\sigma_t$.  For the spatially-constant model, the posterior median spatial lengthscales are roughly 5 miles, with differences attributed to the fact that miles between longitudinal degrees take the formula $\cos(\mbox{deg.~latitude} \times \pi/180) \times 69.17$, i.e., vary with a location's latitude.  We regard this distortion a small price to pay: applying the model using geographic coordinates avoids storing and accessing $O(N^2)$ pairwise distances or recomputing pairwise distances upon every likelihood evaluation.  Indeed, the effective spatial lengthscales for the spatially-varying model vary much more significantly between counties.  With its population density of roughly 74,212 persons per square-mile, Manhattan, NY, exhibits a posterior median effective spatial lengthscale of 0.001 miles, or roughly 5 feet.  With a population density of roughly 35 persons per square-mile, Hancock County, ME, exhibits a posterior median of 2.510 miles.   The spatially-varying model provides similar parameter estimates to the spatially-constant model for most model parameters while providing additional insight into spatial scales of contagion across the urban-rural divide.

\newcommand{\ra}[1]{\renewcommand{\arraystretch}{#1}}
\ra{1.2}
\begin{table}[!t]
	\centering
	\resizebox{1\textwidth}{!}{\begin{tabular}{llll} 
				\toprule
				&& \multicolumn{2}{c}{Posterior median (95\% CrI)}  \\ \cmidrule{3-4}
				Rate component& Parameter & Model 1 (spatially-constant) & Model 2 (spatially-varying)   \\
				\midrule
				Background  & Weight $\mu_0$ & 0.0019 (0.0018, 0.0020)  &  0.0008 (0.0007, 0.0009)  \\
				& Temporal lengthscale $\tau_t$ (wks) & 25.2 (22.3, 29.5) & 51.5 (45.4, 58.0)  \\
				Self-excitatory & Weight $\xi_0$ & 1.001 (0.999, 1.003) & 1.002 (1.000, 1.004) \\
				&	Spatial lengthscale $\sigma_x$ (deg$^*$)  & 0.0798 (0.0794, 0.0802) &  1.479 (1.471, 1.486) \\
				& Temporal lengthscale $\sigma_t$ (wks) & 2.81 (2.79, 2.83) & 2.50 (2.49, 2.52)  \\
				\bottomrule
		\end{tabular}}
		\caption{Posterior medians and 95\% credible intervals for two models' parameters.  Model 1 uses a spatially-constant triggering function \eqref{eq:fixed} and Model 2 uses a spatially-varying triggering function \eqref{eq:varying}.  The unit of measurement for the spatial lengthscale parameter $\sigma_x$ is degrees (lattitude/longitude) for both models, but the effective spatial lengthscale for Model 2 is degrees over (square-root) population density (Table \ref{tab:lengthscales}). 
	}\label{tab:medians}
\end{table}

\begin{table}[!t]
	\centering
	\resizebox{1\textwidth}{!}{\begin{tabular}{lllll} 
				\toprule
				&&& \multicolumn{2}{c}{Spatial lengthscale posterior median (95\% CrI)}  \\ \cmidrule{4-5}
				County & Pop/mi$^2$ & Latitude & Spatially-constant (mi)  & Spatially-varying (mi) \\   \midrule
				Los Angeles County, CA & 2467.79 & 34.20  &  5.05 (5.03, 5.08) & 0.038 (0.038, 0.038) \\  
				Marin County, CA & 504.16 & 38.05  & 4.94 (4.92, 4.97) & 0.182 (0.181,  0.183) \\ 
				Miami-Dade County, FL & 1423.69 & 25.61  &   5.26 (5.23, 5.28) & 0.068 (0.068, 0.069)  \\ 
				Hancock County, ME & 34.96 & 44.56  & 4.74 (4.71, 4.76) & 2.510 (2.497, 2.522) \\  
				Manhattan, NY & 74211.61 & 40.78  & 4.86 (4.83, 4.88) &  0.001 (0.001, 0.001) \\
				Wake County, NC & 1352.23 & 35.79 &   5.01 (4.98, 5.03) & 0.068 (0.069, 0.069)  \\   \bottomrule
		\end{tabular}}
		\caption{Self-excitatory spatial lengthscales for models with spatially-constant \eqref{eq:fixed} and spatially-varying \eqref{eq:varying} triggering kernels are the expected distance between cases and the cases they trigger after $\sigma_t$ weeks.   Translating from units of degrees to miles induces minor distortions that depend on a location's latitude.  Spatially-varying lengthscales adapt to population density.}\label{tab:lengthscales}
	\end{table}

%

\section{Discussion}

We develop a multi-GPU, high-performance computional inference strategy for spatiotemporal Hawkes process models and apply our framework to the analysis of 1 million COVID-19 cases detected in U.S.~hospitals.  Our computational strategy balances within- and between-chain parallelization, assigning individual chains to individual GPU nodes and making specific subsets of GPUs responsible for fine-grained likelihood parallelizations.  We show that this strategy is practical for two different spatiotemporal Hawkes models and pair accelerated Metropolis-Hastings updates with fast spatial location resampling within the Julia programming language. Importantly, both of these updates leave the cut-posterior distribution \eqref{eq:cut1} invariant.  Our spatiotemporal Hawkes process model with spatially-varying kernel provides county specific spatial lengthscales, the posteriors of which provide insight into local spatial contagion dynamics.  Nonetheless, inference for other model parameters appears robust to spatial lengthscale model specification.

This work points to two major lines of future inquiry.  First, how can one scale more complex and flexible spatiotemporal Hawkes process models to massive numbers of observations?   For example, \citet{chiang2022hawkes_} show the benefit of including event specific covariates within their model and allow the self-excitation weight $\xi_0$ to vary through time, interpreting it as an effective reproduction number.  On one hand, these kinds of developments would allow for greater scientific interpretability, allowing one to infer  associations between evolving environmental factors and changing contagion dynamics.  On the other hand, one must avoid significant increases to a model's parameter count when scaling to massive observation counts: more parameters imply a higher-dimensional MCMC state space and the need for longer Markov chains to reach similar effective sample sizes.   Second, tools are needed to accelerate the Hawkes process data scientific workflow in big data contexts.  While our inferential framework fits models relatively quickly, it requires hand-coding individual models in low-level OpenCL. In this manner, model development represents a major bottleneck in the larger data scientific workflow, and this bottleneck becomes all the more limiting if one develops the more complex models discussed above.  A major question is how one can incorporate Hawkes processes and their variants into a probabilistic programming language similar to NIMBLE \citep{de2017programming} or Stan \citep{carpenter2017stan} in a manner that automatically and efficiently maps likelihood computations to device-specific parallel computing resources.

\section*{Acknowledgments}

AJH is supported by grants NIH K25 AI153816, NSF DMS 2152774, and NSF DMS 2236854. MAS is supported by NIH grants U19 AI135995, R01 AI153044 and R01 AI162611. We gratefully acknowledge generous support of Advanced Micro Devices, Inc., including the donation of parallel computing resources used for this research.

\bibliographystyle{sysbio}
\bibliography{refs,refs2}

\appendix

\section{Example OpenCL kernel}\label{sec:kernel}

The following OpenCL kernel computes the log-likelihood \eqref{eqn:model} contribution $\ell_i$ in single precision.  After defining a number of quantities, a GPU uses 256 interleaved threads to  compute memory efficient running sums and maxima for rates $\lambda_{in'}$, using the results as inputs for the log-sum-exp trick.  It then adds the observation's integral term $\Lambda_i$.

\begin{Shaded}
\begin{Highlighting}[numbers=left,fontsize=\footnotesize,]
\PreprocessorTok{\#pragma OPENCL EXTENSION cl\_khr\_fp64 : enable}
\NormalTok{\_\_kernel }\DataTypeTok{void}\NormalTok{ computeLikContribsModel1}\OperatorTok{(}
\NormalTok{                                \_\_global }\AttributeTok{const}\NormalTok{ float2 }\OperatorTok{*}\NormalTok{locations}\OperatorTok{,}
\NormalTok{                                \_\_global }\AttributeTok{const} \DataTypeTok{float} \OperatorTok{*}\NormalTok{times}\OperatorTok{,}
\NormalTok{                                \_\_global }\DataTypeTok{float} \OperatorTok{*}\NormalTok{likContribs}\OperatorTok{,}
                                \AttributeTok{const} \DataTypeTok{float}\NormalTok{ sigmaXprec}\OperatorTok{,}
                                \AttributeTok{const} \DataTypeTok{float}\NormalTok{ tauXprec}\OperatorTok{,}
                                \AttributeTok{const} \DataTypeTok{float}\NormalTok{ tauTprec}\OperatorTok{,}
                                \AttributeTok{const} \DataTypeTok{float}\NormalTok{ omega}\OperatorTok{,}
                                \AttributeTok{const} \DataTypeTok{float}\NormalTok{ theta}\OperatorTok{,}
                                \AttributeTok{const} \DataTypeTok{float}\NormalTok{ mu0}\OperatorTok{,}
                                \AttributeTok{const} \DataTypeTok{int}\NormalTok{ dimX}\OperatorTok{,}
                                \AttributeTok{const} \ExtensionTok{uint}\NormalTok{ locationStart}\OperatorTok{,}
                                \AttributeTok{const} \ExtensionTok{uint}\NormalTok{ locationEnd}\OperatorTok{,}
                                \AttributeTok{const} \ExtensionTok{uint}\NormalTok{ locationCount}\OperatorTok{)} \OperatorTok{\{}
    \AttributeTok{const} \ExtensionTok{uint}\NormalTok{ i }\OperatorTok{=}\NormalTok{ locationStart }\OperatorTok{+}\NormalTok{ get\_group\_id}\OperatorTok{(}\DecValTok{0}\OperatorTok{);}
    \AttributeTok{const} \ExtensionTok{uint}\NormalTok{ gid }\OperatorTok{=}\NormalTok{ get\_group\_id}\OperatorTok{(}\DecValTok{0}\OperatorTok{);}
    \AttributeTok{const} \ExtensionTok{uint}\NormalTok{ lid }\OperatorTok{=}\NormalTok{ get\_local\_id}\OperatorTok{(}\DecValTok{0}\OperatorTok{);}
    \ExtensionTok{uint}\NormalTok{ j }\OperatorTok{=}\NormalTok{ get\_local\_id}\OperatorTok{(}\DecValTok{0}\OperatorTok{);}
\NormalTok{    \_\_local }\DataTypeTok{float}\NormalTok{ scratch}\OperatorTok{[}\DecValTok{256}\OperatorTok{];}
\NormalTok{    \_\_local }\DataTypeTok{float}\NormalTok{ scratch2}\OperatorTok{[}\DecValTok{256}\OperatorTok{];}
    \AttributeTok{const}\NormalTok{ float2 vectorI }\OperatorTok{=}\NormalTok{ locations}\OperatorTok{[}\NormalTok{i}\OperatorTok{];}
    \AttributeTok{const} \DataTypeTok{float}\NormalTok{ timeI }\OperatorTok{=}\NormalTok{ times}\OperatorTok{[}\NormalTok{i}\OperatorTok{];}
    \DataTypeTok{float}\NormalTok{        sum }\OperatorTok{=} \FloatTok{0.0}\BuiltInTok{f}\OperatorTok{;}
    \DataTypeTok{float}\NormalTok{ mu0TauXprecDTauTprec }\OperatorTok{=}\NormalTok{ mu0 }\OperatorTok{*}\NormalTok{ pow}\OperatorTok{(}\NormalTok{tauXprec}\OperatorTok{,}\NormalTok{dimX}\OperatorTok{)} \OperatorTok{*}\NormalTok{ tauTprec}\OperatorTok{;}
    \DataTypeTok{float}\NormalTok{ thetaSigmaXprecDOmega }\OperatorTok{=}\NormalTok{ theta }\OperatorTok{*}\NormalTok{ pow}\OperatorTok{(}\NormalTok{sigmaXprec}\OperatorTok{,}\NormalTok{dimX}\OperatorTok{)} \OperatorTok{*}\NormalTok{ omega}\OperatorTok{;}
    \ControlFlowTok{while} \OperatorTok{(}\NormalTok{j }\OperatorTok{\textless{}}\NormalTok{ locationCount}\OperatorTok{)} \OperatorTok{\{}
        \AttributeTok{const} \DataTypeTok{float}\NormalTok{ timDiff }\OperatorTok{=}\NormalTok{ timeI }\OperatorTok{{-}}\NormalTok{ times}\OperatorTok{[}\NormalTok{j}\OperatorTok{];}
        \AttributeTok{const}\NormalTok{ float2 vectorJ }\OperatorTok{=}\NormalTok{ locations}\OperatorTok{[}\NormalTok{j}\OperatorTok{];}
        \AttributeTok{const}\NormalTok{ float2 difference }\OperatorTok{=}\NormalTok{ vectorI }\OperatorTok{{-}}\NormalTok{ vectorJ}\OperatorTok{;}

        \AttributeTok{const} \DataTypeTok{float}\NormalTok{ distancesq }\OperatorTok{=}\NormalTok{ dot}\OperatorTok{(}\NormalTok{difference}\OperatorTok{.}\NormalTok{lo}\OperatorTok{,}\NormalTok{ difference}\OperatorTok{.}\NormalTok{lo}\OperatorTok{)} \OperatorTok{+}
\NormalTok{                            dot}\OperatorTok{(}\NormalTok{difference}\OperatorTok{.}\NormalTok{hi}\OperatorTok{,}\NormalTok{ difference}\OperatorTok{.}\NormalTok{hi}\OperatorTok{);}

        \AttributeTok{const} \DataTypeTok{float}\NormalTok{ innerContrib }\OperatorTok{=}\NormalTok{ mu0TauXprecDTauTprec }\OperatorTok{*}
                            \OperatorTok{((}\NormalTok{timDiff }\OperatorTok{!=} \DecValTok{0}\OperatorTok{)} \OperatorTok{?}\NormalTok{ pdf}\OperatorTok{(}\NormalTok{timDiff }\OperatorTok{*}\NormalTok{ tauTprec}\OperatorTok{)} \OperatorTok{:} \FloatTok{0.0}\BuiltInTok{f}\OperatorTok{)} \OperatorTok{+}
\NormalTok{                            thetaSigmaXprecDOmega }\OperatorTok{*}
                            \OperatorTok{((}\NormalTok{timDiff }\OperatorTok{\textgreater{}} \DecValTok{0}\OperatorTok{)} \OperatorTok{?}\NormalTok{ exppdf\_2sq}\OperatorTok{(}\NormalTok{omega }\OperatorTok{*}\NormalTok{ timDiff}\OperatorTok{,} 
\NormalTok{                            distancesq }\OperatorTok{*}\NormalTok{ sigmaXprec }\OperatorTok{*}\NormalTok{ sigmaXprec}\OperatorTok{)} \OperatorTok{:} \FloatTok{0.0}\BuiltInTok{f}\OperatorTok{);}
\NormalTok{        sum }\OperatorTok{+=}\NormalTok{ innerContrib}\OperatorTok{;}
\NormalTok{        j }\OperatorTok{+=} \DecValTok{256}\OperatorTok{;}
    \OperatorTok{\}}

\NormalTok{    scratch}\OperatorTok{[}\NormalTok{lid}\OperatorTok{]} \OperatorTok{=}\NormalTok{ sum}\OperatorTok{;}
\NormalTok{    scratch2}\OperatorTok{[}\NormalTok{lid}\OperatorTok{]} \OperatorTok{=}\NormalTok{ sum}\OperatorTok{;}
    \CommentTok{// find maximum}
\NormalTok{    barrier}\OperatorTok{(}\NormalTok{CLK\_LOCAL\_MEM\_FENCE}\OperatorTok{);}
    \ControlFlowTok{for}\OperatorTok{(}\DataTypeTok{int}\NormalTok{ k }\OperatorTok{=} \DecValTok{1}\OperatorTok{;}\NormalTok{ k }\OperatorTok{\textless{}} \DecValTok{256}\OperatorTok{;}\NormalTok{ k }\OperatorTok{\textless{}\textless{}=} \DecValTok{1}\OperatorTok{)} \OperatorTok{\{}
\NormalTok{        barrier}\OperatorTok{(}\NormalTok{CLK\_LOCAL\_MEM\_FENCE}\OperatorTok{);}
        \ExtensionTok{uint}\NormalTok{ mask }\OperatorTok{=} \OperatorTok{(}\NormalTok{k }\OperatorTok{\textless{}\textless{}} \DecValTok{1}\OperatorTok{)} \OperatorTok{{-}} \DecValTok{1}\OperatorTok{;}
        \ControlFlowTok{if} \OperatorTok{((}\NormalTok{lid }\OperatorTok{\&}\NormalTok{ mask}\OperatorTok{)} \OperatorTok{==} \DecValTok{0}\OperatorTok{)} \OperatorTok{\{}
            \ControlFlowTok{if} \OperatorTok{(}\NormalTok{scratch2}\OperatorTok{[}\NormalTok{lid}\OperatorTok{]} \OperatorTok{\textless{}}\NormalTok{ scratch2}\OperatorTok{[}\NormalTok{lid }\OperatorTok{+}\NormalTok{ k}\OperatorTok{])}
\NormalTok{                scratch2}\OperatorTok{[}\NormalTok{lid}\OperatorTok{]} \OperatorTok{=}\NormalTok{ scratch2}\OperatorTok{[}\NormalTok{lid }\OperatorTok{+}\NormalTok{ k}\OperatorTok{];}
        \OperatorTok{\}}
    \OperatorTok{\}}

\NormalTok{    barrier}\OperatorTok{(}\NormalTok{CLK\_LOCAL\_MEM\_FENCE}\OperatorTok{);}

    \CommentTok{// log{-}sum{-}exp trick, find maximum and normalize. }
    \DataTypeTok{float}\NormalTok{ maximum }\OperatorTok{=}\NormalTok{ max}\OperatorTok{(}\NormalTok{scratch2}\OperatorTok{[}\DecValTok{0}\OperatorTok{],} \OperatorTok{(}\DataTypeTok{float}\OperatorTok{)} \FloatTok{1e{-}40}\OperatorTok{);}
\NormalTok{    scratch}\OperatorTok{[}\NormalTok{lid}\OperatorTok{]} \OperatorTok{=}\NormalTok{ scratch}\OperatorTok{[}\NormalTok{lid}\OperatorTok{]} \OperatorTok{/}\NormalTok{ maximum}\OperatorTok{;}
\NormalTok{    barrier}\OperatorTok{(}\NormalTok{CLK\_LOCAL\_MEM\_FENCE}\OperatorTok{);}
    \ControlFlowTok{for}\OperatorTok{(}\DataTypeTok{int}\NormalTok{ k }\OperatorTok{=} \DecValTok{1}\OperatorTok{;}\NormalTok{ k }\OperatorTok{\textless{}} \DecValTok{256}\OperatorTok{;}\NormalTok{ k }\OperatorTok{\textless{}\textless{}=} \DecValTok{1}\OperatorTok{)} \OperatorTok{\{}
\NormalTok{        barrier}\OperatorTok{(}\NormalTok{CLK\_LOCAL\_MEM\_FENCE}\OperatorTok{);}
        \ExtensionTok{uint}\NormalTok{ mask }\OperatorTok{=} \OperatorTok{(}\NormalTok{k }\OperatorTok{\textless{}\textless{}} \DecValTok{1}\OperatorTok{)} \OperatorTok{{-}} \DecValTok{1}\OperatorTok{;}
        \ControlFlowTok{if} \OperatorTok{((}\NormalTok{lid }\OperatorTok{\&}\NormalTok{ mask}\OperatorTok{)} \OperatorTok{==} \DecValTok{0}\OperatorTok{)} \OperatorTok{\{}
\NormalTok{            scratch}\OperatorTok{[}\NormalTok{lid}\OperatorTok{]} \OperatorTok{+=}\NormalTok{ scratch}\OperatorTok{[}\NormalTok{lid }\OperatorTok{+}\NormalTok{ k}\OperatorTok{];}
        \OperatorTok{\}}
    \OperatorTok{\}}
\NormalTok{    barrier}\OperatorTok{(}\NormalTok{CLK\_LOCAL\_MEM\_FENCE}\OperatorTok{);}
\NormalTok{    scratch}\OperatorTok{[}\DecValTok{0}\OperatorTok{]} \OperatorTok{=}\NormalTok{ max}\OperatorTok{(}\NormalTok{scratch}\OperatorTok{[}\DecValTok{0}\OperatorTok{],} \OperatorTok{(}\DataTypeTok{float}\OperatorTok{)} \FloatTok{1e{-}40}\OperatorTok{);}

\NormalTok{    barrier}\OperatorTok{(}\NormalTok{CLK\_LOCAL\_MEM\_FENCE}\OperatorTok{);}
    \ControlFlowTok{if} \OperatorTok{(}\NormalTok{lid }\OperatorTok{==} \DecValTok{0}\OperatorTok{)} \OperatorTok{\{}
\NormalTok{        likContribs}\OperatorTok{[}\NormalTok{gid}\OperatorTok{]} \OperatorTok{=}\NormalTok{ log}\OperatorTok{(}\NormalTok{maximum}\OperatorTok{)} \OperatorTok{+}\NormalTok{ log}\OperatorTok{(}\NormalTok{scratch}\OperatorTok{[}\DecValTok{0}\OperatorTok{])} \OperatorTok{+}\NormalTok{ theta }\OperatorTok{*}
        \OperatorTok{(}\NormalTok{ exp}\OperatorTok{({-}}\NormalTok{omega}\OperatorTok{*(}\NormalTok{times}\OperatorTok{[}\NormalTok{locationCount}\OperatorTok{{-}}\DecValTok{1}\OperatorTok{]{-}}\NormalTok{times}\OperatorTok{[}\NormalTok{i}\OperatorTok{])){-}}\DecValTok{1} \OperatorTok{)} \OperatorTok{{-}} 
\NormalTok{        mu0 }\OperatorTok{*} \OperatorTok{(}\NormalTok{ cdf}\OperatorTok{((}\NormalTok{times}\OperatorTok{[}\NormalTok{locationCount}\OperatorTok{{-}}\DecValTok{1}\OperatorTok{]{-}}\NormalTok{times}\OperatorTok{[}\NormalTok{i}\OperatorTok{])*}\NormalTok{tauTprec}\OperatorTok{){-}}
\NormalTok{        cdf}\OperatorTok{({-}}\NormalTok{times}\OperatorTok{[}\NormalTok{i}\OperatorTok{]*}\NormalTok{tauTprec}\OperatorTok{)} \OperatorTok{)} \OperatorTok{;}
    \OperatorTok{\}}
\OperatorTok{\}}
\end{Highlighting}
\end{Shaded}

%
%
%
%

\end{document}